\documentclass[aps,tightenlines,eqsecnum,floats,amsmath,amssymb,nofootinbib,prd,showpacs,twocolumn]{revtex4}
\usepackage{amsmath}
\usepackage{amsfonts}
\usepackage{graphicx}
\newcommand{\be}{\nopagebreak[3]\begin{equation}}
\newcommand{\ee}{\end{equation}}
\newcommand{\ba}{\nopagebreak[3]\begin{eqnarray}}
\newcommand{\ea}{\end{eqnarray}}

\newcommand{\bmult}{\nopagebreak[3]\begin{multline}}
\newcommand{\emult}{\end{multline}}

\begin{document}
\title{Dynamical decoupling with tailored waveplates for long distance communication using polarization qubits}
\author{Bhaskar Roy Bardhan}
\email{broyba1@lsu.edu}
\author{Katherine L.\ Brown}
\author{Jonathan P.\ Dowling}
\affiliation{Department of Physics and Astronomy and Hearne Institute of Theoretical Physics, Louisiana State University, Baton Rouge, LA 70803}

\date{\today }

\begin{abstract}

We address the issue of dephasing effects in flying polarization qubits propagating through optical fiber by using the method of dynamical decoupling. The control pulses are implemented with half waveplates suitably placed along the realistic lengths of the single mode optical fiber. The effects of the finite widths of the waveplates on the polarization rotation are modeled using tailored refractive index profiles inside the waveplates. We show that dynamical decoupling is effective in preserving the input qubit state with the fidelity close to one when the polarization qubit is subject to the random birefringent noise in the fiber, as well the rotational imperfections (flip-angle errors) due to the finite width of the waveplates.


\end{abstract}


\maketitle

\section{Introduction}

Quantum information processing (QIP) has gained huge interest  over the last few decades. This is because of the fact that it is is potentially able to solve many problems faster than the classical counterpart as well as provide secure communication channel. However, the inevitable interaction of the qubits with a noisy environment causes the loss of coherence leading to errors in the processing of the quantum information. This effect is known as decoherence and it limits the time scale over which quantum information can be retained and the distance over which it can be transmitted.


Among the various strategies developed to combat the decoherence effects, we consider the method called dynamical decoupling (DD)~\cite{BB,Lidar,Uhrig,Viola,ViolaKnill,Uhrig2009,West,Kuo}. This is a relatively simple but effective technique, which uses sequences of  external control pulses applied to the system qubits to reduce (or average out) the interaction of the system with the environment. A significant advantage of the DD techniques are that they do not require any ancilla qubits or encoding or measurement overheads.

Most of the theoretical works on DD have hitherto considered only the case of ideal pulses. This means that the pulses were assumed to be instantaneous and infinitely strong in the sense of a $\delta$-shaped pulse. In that case, we can ignore the effect of the noise-inducing environment during the application of the pulses. However, in any realistic physical implementation, this is no longer the case as the pulses generally have a finite duration ($\tau_{\textrm p}$). As the effects of the applied pulses can be viewed as rotations of the Bloch vector on the unit sphere, the physical pulses result in rotational imperfections. Such imperfections typically have a cumulative effect, leading to progressive unaccounted phase errors in the echo signal. Such an effect leads to a considerable reduction of the performance of the DD sequences.

Here, we extend the idea of dynamical decoupling with realistic pulses to long distance communication using single photons. In optical quantum information processing, information is usually encoded in the polarization state of photons (which is called a polarization qubit) and photons are then typically routed through optical fibers or waveguides. Transmission of the photons through such optical fibers can be helpful in using them as optical quantum memory~\cite{Gingrich,quantummemory}, distributed quantum computation~\cite{distributednetworks} and quantum cryptography~\cite{Bouwmeester}. For the last few decades, quantum communication using propagation of the single photons through optical fibers has therefore emerged as a very active area of research and commercial development. The initial state of a polarization qubit can usually be very well prepared while it is much more challenging to preserve this state along the communication channel before it reaches the final detection stage.
 
As the single photon propagates through the fiber, external effects such as temperature, stress, etc., within the fiber can randomly affect the polarization state of the photons~\cite{Ulrich}. This type of noise is referred to as birefringent noise as these uncontrollable fluctuating factors cause the birefringence, i.e. the refractive index difference $\Delta n=|n_{H}-n_{V}|$ (where $n_{H}$ and  $n_{V}$ denote the refractive indices corresponding to the horizontal and vertical polarizations, respectively) along the fiber to change randomly. The effect results in the polarization state of the single photons to change very rapidly, making it impractical to correct for it by calibration. In practice, optical fibers used for communications with light can be several hundreds of kilometers long and the birefringence in such long optical fibers can totally destroy the information stored in the polarization qubits. 

It is therefore crucial to protect the flying polarization qubits against such detrimental dephasing effects induced by the fiber. Several recent studies have looked at suppression of these effects on the polarization qubits. Wu and Lidar~\cite{Wu} showed that dynamical decoupling could be applied  for reducing quantum noise in optical fiber. Massar and Popescu presented a method to reduce polarization mode dispersion in optical fibers using controlled polarization rotations~\cite{Popescu}. In our previous work~\cite{BRB}, we numerically simulated random birefringent noise along realistic fiber length and showed that application of DD could well preserve the input polarization qubit from such noise. We chose the CPMG sequence~\cite{CPMG} of DD because this sequence has been shown to be robust against a variety of dephasing and control pulse errors~\cite{Freeman,Morton,Ajoy}. We restricted our analysis to ideal pulses implemented with zero-width half-waveplates along the fiber.

In this paper, we investigate in details the issue of polarization dephasing by using $\emph{finite width}$ waveplates which are likely to cause some additional errors, apart from the random birefringent dephasing noise. Finite widths of the birefringent waveplates directly affect the phase of the photon transmitted through the waveplates leading to further loss of coherence. Introducing tailored refractive index profiles within the waveplates, we show that it is possible to address such detrimental effects with DD techniques when implemented with waveplates at the prescribed locations along the fiber. Estimates of the required inter-waveplate distance are provided for each of the refractive index profiles, and this information will be enough for an experimenter to know to successfully implement the faithful long-distance quantum communication channel in practice. 


This paper is organized as follows. In Sec II, we briefly introduce the basic ideas of DD (with ideal and realistic pulses). We discuss the nature of birefringent noise in optical fiber and how the DD techniques can be applied to table such noise with finite-width waveplates in Sec. III. Section IV contains our numerical results and comparative analysis for various tailored refractive index profiles. Finally, we conclude with a brief summary of the results with accompanying discussions.

%


\section{Dynamical Decoupling}

\subsection{General Idea}

Dynamical decoupling is an effective method to time-reverse the system-bath interaction by applying sequences of sufficiently fast and strong pulses. As a result of application of the sequence of pulses, the interaction of a qubit system with the environment is reduced, thus retaining the quantum information for longer time (or distance). The most general Hamiltonian describing the evolution of a system coupled to a bath can be written as
\begin{equation}
\label{totalhamiltonian}
H_{\rm tot}=H_{\rm S}\otimes I_{\rm B}+I_{\rm S}\otimes H_{\rm B}+H_{\rm I},
\end{equation}
where $H_S$ and $H_B$ are the system and bath Hamiltonians, respectively.  Since it is difficult to control the states of the environment, the control pulses need to act on the system, and the effect of these pulses are described as a refocusing of the system-environment interaction by a control Hamiltonian $H_{C}(t)$. When the system is a qubit undergoing dephasing, we can write the interaction Hamiltonian $H_{\rm I}$ as
\begin{equation}
H_{\rm I} = \sigma_{z} \otimes B_{Z},
\end{equation}
where $\sigma_{z}$ is the Pauli $Z$ spin operator and $B_{Z}$ is a bath operator which couples to the photonic qubit, causing dephasing.

\begin{figure}
\centering
		\includegraphics[width=\columnwidth]{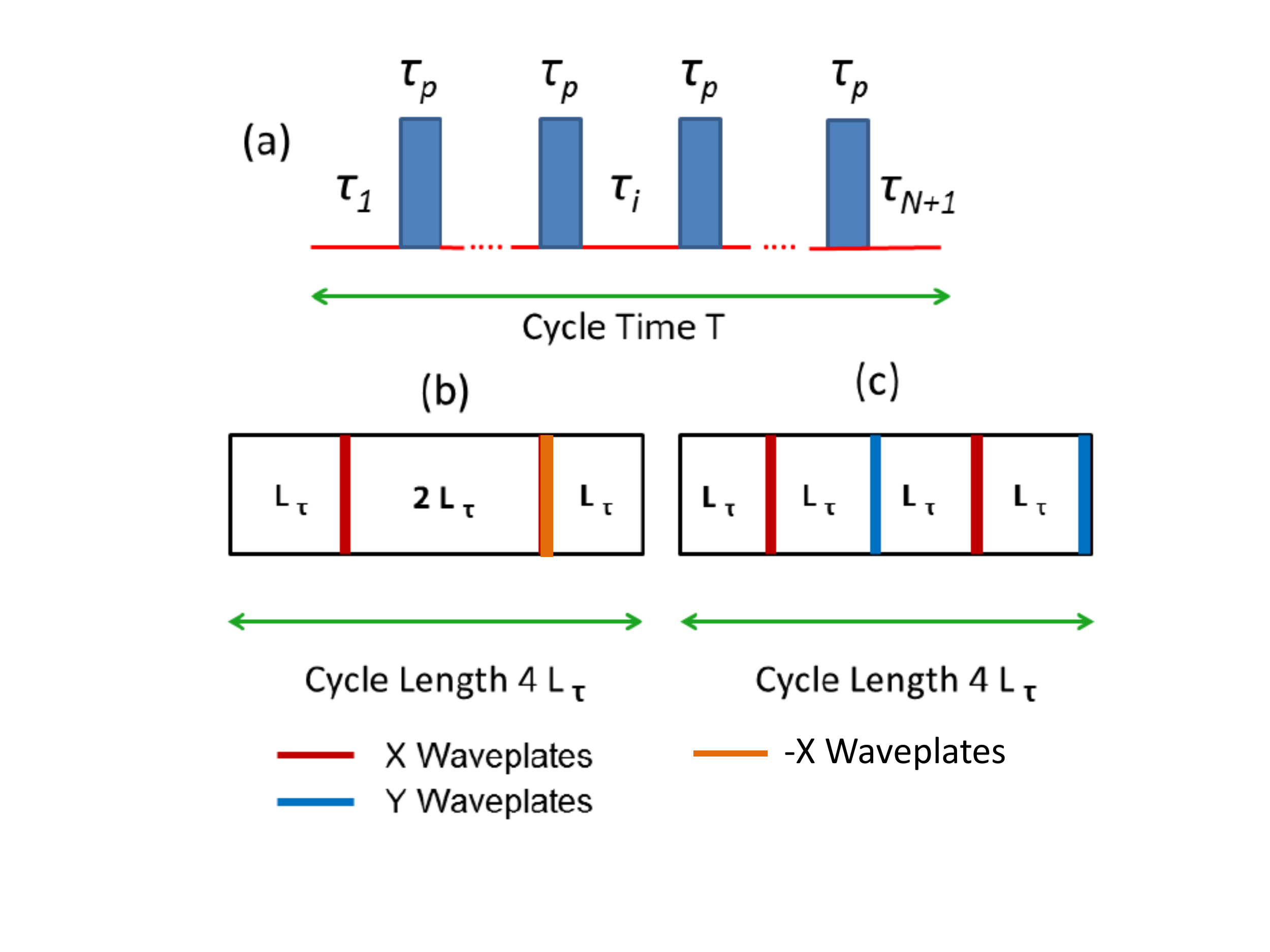}
	\caption{(Color Online) (a) Schematic representation of a cycle of general dynamical sequence with pulses of duration $\tau_{p}$, (b) CPMG sequence with two pulses in a cycle, and (c) XY-4 sequence with alternated X and Y waveplates. Here X, -X, an Y waveplates implement the $\pi$-rotations around $x, -x,$ and $y$ axes, respectively.}
\end{figure}

Let us now consider a single cycle of a DD sequence having a period $T$. The evolution operator describing the evolution of the total system from $0$ to $T$ (Fig. 1), in the rotating frame, can be written as~\cite{AlvarezPRA}
\begin{equation}
\label{tothamiltonian}
U(T)=U_{f}(\tau_{N+1})\prod_{i=1}^N U_{C}^{i}(\tau_{p})U_{f}(\tau_{i}),
\end{equation}
where the free evolution operator is (Eq.~(\ref{totalhamiltonian}) given by
\begin{equation}
U_{\rm f}(\tau)=\exp[-iH_{\rm tot}(\tau)]
\end{equation}
and evolution operator acting during the application of pulses is
\begin{equation}
\label{Uc}
U_{C}^{i}(\tau_{p})=\mathcal{T}\exp\left[-i\int_{0}^{\tau_{p}} dt^{'}(H_{tot}+H_{C}^{i}(t^{'})\right]
\end{equation}
Here $\mathcal{T}$ is the standard time-ordering operator and Hamiltonian of the control pulse can be written as
\begin{equation}
H_{C}(t)=\vec \sigma. \vec f(t)
\end{equation}
where $\vec f(t)=(f_{x}(t), f_{y}(t), f_{z}(t))$ is the vector defining the shape of the pulse and $\vec \sigma$ is the vector of the Pauli matrices. The axis of rotation due to the applied pulse at the instant $t$ is given by the unit vector $\frac{\vec f(t)}{|\vec f(t)|}$ which implies that rotation due to the pulse can be adjusted by tuning the pulse shapes. The underlying principle of dynamical decoupling is to select a ``pulse sequence'' $f(t)$ which causes the integrated time evolution of the interaction Hamiltonian to coherently average to zero.

Since the evolution described in Eq.~\ref{tothamiltonian} is necessarily  an unitary one, it can be written as the exponential of a hermitian operator $H_{\rm eff}$
\begin{equation}
U(T)=\exp[-i H_{\rm eff} T],
\end{equation}
where $H_{\rm eff}$ can be written as a series expansion using the average Hamiltonian theory~\cite{Haeberlen2}
\begin{equation}
\label{avghamiltonian}
H_{\rm eff}=H^{(0)}+H^{(1)}+.......=\sum_{n=0}^{\infty} H^{(n)}.
\end{equation}

An ideal DD sequence, i.e. a DD sequence with instantaneous pulses, makes $H^{(0)}=H_{\rm B}$ in general and better performance of a DD sequence usually corresponds to progressively eliminating the higher order terms in such an expansion~\cite{Liu,Biercuk,Lidar} (e.g. Magnus expansion~\cite{Blanes}). Prominent examples of DD schemes are the periodic DD (PDD)~\cite{Viola}, Carr-Purcell DD (CP)~\cite{CP}, Carr-Purcell-Meiboom-Gill (CPMG)~\cite{CPMG}, concatenated DD (CDD)~\cite{Lidar} and Uhrig DD (UDD)~\cite{Uhrig}

\subsection{Ideal and Real Pulses}

Ideally, DD pulses are assumed to be strong and instantaneous pulses applied fast enough compared to the internal dynamics of the environment. Under the assumption of weak coupling to the environment, the evolution operator in Eq.~\ref{Uc} (in the rotating frame) simplifies to
\begin{equation}
U_{C}^{i}(\tau_{p})=\exp[-i \sigma_{\alpha} \theta_{p}/2].
\end{equation}
Here $\alpha=x, y, z$ and $\theta_{p}=\omega_{p} \tau_{p}$ ($\omega_{p}$ being the frequency of the pulses) is the rotation around the $\alpha$ axis. For ideal instantaneous pulses which implement $\pi$ rotations, the angle $\theta_{p}$ will be $\pi$. 

However, imperfect pulses can result in errors in the rotation axis as well as angle of rotation. We can write the resulting propagator as the product of the ideal pulse propagator and a rotational error $\exp[-i \sigma_{{e}_{i}} \theta_{{e}_{i}}/2]$~\cite{AlvarezPRA}, due to the $i$-th waveplate,

\begin{equation}
\label{Uc_error}
U_{C}^{i}(\tau_{p})=\exp[-i \sigma_{{e}_{i}} \theta_{{e}_{i}}/2] \exp[-i \sigma_{\alpha} \theta_{p}/2].
\end{equation}

The modified free evolution operator in presence of the pulse errors can be written as
\begin{equation}
{U_{f}} (\tau_{i}, \tau_{p})=U_{f}(\tau_{i}) \exp[-i \sigma_{{e}_{i}} \theta_{{e}_{i}}/2].
\end{equation}

The total evolution operation from Eq.~(\ref{tothamiltonian}) then reads
\begin{equation}
\label{finpropagator}
U(T)=U_{f_{N+1}}^\prime (\tau_{N+1}, \tau_{p})\prod_{i=1}^N U_{C}^{i}(0)U_{f}^\prime (\tau_{i}, \tau_{p}).
\end{equation}

A good choice of DD sequence should make $U(T) \approx \mathbb{I}$ (the identity) in the presence of the pulse errors defined above, preserving the initial qubit against decoherence along with rotational imperfections. Khodjasteh and Lidar analyzed the cumulative effects in pulse sequences and provided an optimum pulse interval for realistic pulses with a fixed minimal pulse width $\tau_{\textrm p,min}$~\cite{Lidar2007}. Uhrig and Pasini showed the optimized performance of the DD sequences for considering realistic control pulses of finite duration and amplitude~\cite{Pasini2010,Pasini2010NJP}. Composite pulse sequences such as BB1, CORPSE, and SCORPSE, have been shown to correct systematic pulse errors (which might include pulse amplitude, phase and frequency errors)~\cite{Brown,Cummins1,Cummins2,Mottonen,Suter2011}. Pulse shaping is another method that is used to counteract environmental noise effects during the finite duration of the real pulses~\cite{Pasini2008,Pryadko2008,Mottonen,Sengupta2005}.

\begin{figure}[b!]
\centering
		\includegraphics[height=\columnwidth,width=3.0in,angle=-90]{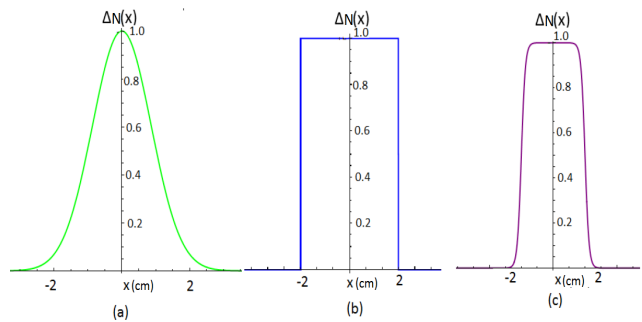}
	\caption{(Color Online) Refractive index profiles $\Delta N(x)$ generating the phase error, where $x$ represents the distance within the waveplate: (a) Gaussian, (b) Rectangular, and (c) Hyperbolic tangent [as defined in the text and in Eq. (4.2)].}
\end{figure}

\section{Suppressing Birefringent Dephasing with DD}

\subsection{Effect of birefringent dephasing on the polarization qubits}

In a long distance communication channel (typically of lengths 10-1000 km), often made with polarization-preserving birefringent fiber, polarization qubits are likely to experience random effects due to changes in temperature, stress, etc. during propagation. The characteristic length scales for such changes may be several meters, i.e. lengths smaller than the fiber beat lengths~\cite{Galtarossa,Popescu}. We approximate the communication channel as a continuously connected fiber elements which have sections of constant birefringence on the order of this length scale. In the following, we consider the evolution of pure single photon state through such channel.

Any arbitrary  single-photon polarization state can be characterized by its polarization and frequency spectrum as~\cite{Berglund}
\begin{equation}
\left|\psi(0)\right\rangle=(\alpha |H\rangle+\beta |V\rangle)\otimes \int d\omega \epsilon(\omega)|\omega\rangle,
\end{equation}
with $|H\rangle$ and $|V\rangle$ denoting the horizontal and vertical polarization states having arbitrary complex amplitudes $\alpha$ and $\beta$, respectively (which satisfy $|\alpha|^{2}+|\beta|^{2}=1$) and $\epsilon(\omega)$ being the complex amplitude corresponding to the frequency $\omega$. If the input photon is transmitted freely for a length $L$ through the birefringent fiber, then the qubit state becomes
\begin{eqnarray}
\left|\psi(L)\right\rangle &=& \nonumber \alpha |H\rangle \otimes \int d \omega \epsilon(\omega) \exp(i\omega n_{H}L/c)|\omega\rangle \\ && \nonumber +\beta |V\rangle \otimes \int d \omega \epsilon(\omega) \exp(i\omega n_{V}L/c)|\omega\rangle \\
\end{eqnarray}
Here  $\Delta n(x)=|n_{H} (x)-n_{V} (x)|$ is the birefringence in the fiber that changes along the distance $x$ in the fiber, causing the dephasing in the single photon polarization state. The phase accumulated by the qubit for the length $L$ is given by
 \begin{eqnarray}
 \triangle \phi &=& \phi_{H}-\phi_{V} = \frac{\omega}{c}\int_{0}^{L}\Delta n(x) dx \nonumber \\ && =(2\pi/\lambda) \int_{0}^{L}\Delta n(x) dx.
 \end{eqnarray}

The final polarization state obtained is
\begin{equation}
\rho(L)=\begin{bmatrix}
  |\alpha|^{2} &  \alpha \beta^{*}F^{*}(L)  \\
 \alpha^{*}\beta F(L) & |\beta|^{2}
   \end{bmatrix}
\end{equation}
where the decoherence function, $F(L)=\int d\omega  |\epsilon(\omega)|^{2} \exp(i\omega \Delta n L/c)$, gives the decay of the off-diagonal elements as the qubit propagates through the fiber. This analysis indicates that the interaction of a polarization qubit with the fiber during its propagation causes the qubit to lose coherence. As a result, the phase of the qubit becomes randomized and the quantum information stored in it is eventually lost.

\subsection{Characterizing the dephasing noise}

If we describe the fluctuations in the fiber to be random, stochastic fluctuation of the refractive index difference $\Delta n$ can be simulated as a Gaussian-distributed zero mean random process. In this case, the noise is completely defined by the first order correlation function at two points $x_{1}$ and $x_{2}$ inside the fiber given by

\begin{equation}
\label{correlfun}
\langle \triangle n(x_{1})\triangle n(x_{2})\rangle=\exp{\left[-\triangle n^{2}/2\sigma_{\triangle n}^{2}\right]}
\end{equation}

Here $\langle \ldots \rangle$ represents the stochastic average, i.e. with respect to the realizations of the birefringent noise and $\vert x_{1}-x_{2}\vert$ is considered to be less than correlation length. (Estimates of the correlation lengths for a typical optical fiber are given in Ref.~\cite{Galtarossa}.).

The Fourier transform $S(k)$ of the correlation function in Eq.~(\ref{correlfun}) is $S(k)=\exp{\left[-\frac{k^{2} \sigma_{n}^{2}}{2}\right]}$, which is related to the decoherence function (defined as the ratio of the final and initial off-diagonal terms of the density matrix)~\cite{Cywinski,BRB}. In our previous work, we showed that by using the decoherence function formalism it is possible to reduce the dephasing rate with the coherence preserved for a longer length along the fiber with a suitable DD sequence.


\subsection{Choice of DD to preserve the polarization qubit}

The facts that the DD pulses are implemented with waveplates in our scheme, and the axis of rotation is fixed by the orientation of the optical axis of the birefringent waveplates, highly restrict our choice of DD methods. It is technically formidable task to have precise control of the varied orientations of the optic axes as required in each sequence of the composite pulse sequences such as BB1, CORPSE, SCORPSE, or KDD (although all of them generally provide robust performance against pulse errors). Moreover, these sequences typically require large number of pulses in each cycle, which in our case of long-distance fibers affects scalability. 

In our case, we have to choose DD sequences which provide robust performance in presence of our dephasing model, which has randomly distributed noise with Gaussian spectral density as specified in Eq.(~\ref{correlfun}). An optimized sequence such as UDD is also not a good choice as UDD works best when the noise has a sharp high-frequency cut-off~\cite{Ajoy,Bollinger,Jenista,Du}). In the case of UDD, with single or multi-axis control, pulse errors generally accumulate with higher orders. A few recent studies also indicate that high-order UDD or concatenated DD sequences in general lose their advantage when the pulse intervals are strongly constrained~\cite{Alvarez2012,Pasini2010}. 


\emph{CPMG and XY-4 Sequences}:  In order to preserve polarization states in a fiber, the CPMG sequence has been shown to work best in such Gaussian-distributed random birefringent noise~\cite{Ajoy,Cywinski,BRB}. Another motivation for using CPMG is that this sequence is extremely robust against all pulse imperfections, when used with the longitudinal states, while giving marginally better results to preserve the transverse components of polarization~\cite{Suter2011,Freeman,Morton,AlvarezPRA,Witzel,Yao,Ajoy,BRB}. It requires $\pi$ rotations around a fixed axis which can be easily set by  orienting the optical axis within half-waveplates.

For very similar reasons, we then consider the XY-4 sequence (which requires alternating $\pi$ rotations around x and y axes), and is known to provide excellent performance in presence of pulse errors by preserving both the longitudinal and transverse components of polarization~\cite{Suter2011,AlvarezPRA,Maudsley,Gullion}. These sequences act as high-pass filters that effectively filter out the components of the $H_{\rm I}$ which vary slowly compared to $\tau$. In both sequences, the total evolution operator after one cycle, defined in Eq.~(\ref{finpropagator}), is $U_{T} \approx \mathbb{I}+\mathcal{O}(\omega_{c}^{2} \tau^{2})$ ($\tau_{c}=1/ \omega_{c}$ is the correlation time of the environment). Hence, the errors (due to both dephasing and pulse imperfections) resulting in the randomization in the phase of the polarization state coming out of the fiber at the end, can be reduced with CPMG and XY-4 upto the first order in $\omega_{c} \tau$ for each cycle.


\section{Numerical simulations and Results}

\begin{figure}[t!]
\centering
		\includegraphics[height=\columnwidth,angle=-90]{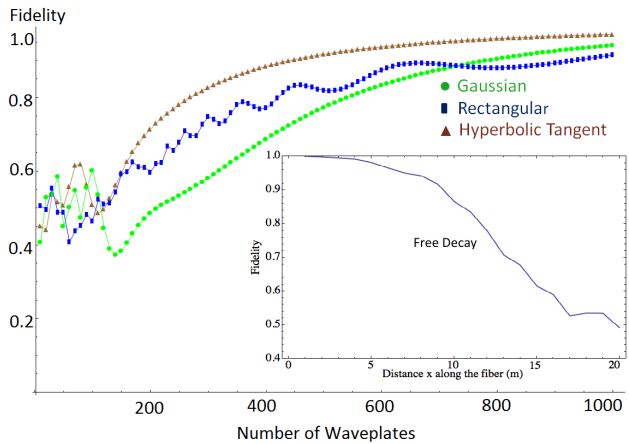}
	\caption{(Color Online) Fidelity obtained with XY-4 waveplates is shown with variation of the number of waveplates for different refractive index profiles. Total length of the optical fiber is 10 km, and number of randomly generated phase profiles to obtain the average fidelity is 500. Inset: Fidelity decay (free decay) without the waveplates in the fiber.}
\end{figure}

We focus only on the dephasing noise (i.e. the noise represented by the Pauli $Z$ operator), and neglect the errors due to energy dissipation inside the fiber. This assumption is justified since we restrict our calculations for the wavelengths in the telecommunication band (around 1550 nm) where the optical losses are minimal~\cite{Hiskett,Hadfield}. In fact, any state prepared along a direction that is different from the $z$-axis will undergo dephasing, i.e. a rotation around the z-axis, due to the birefringent noise along the fiber.

\emph{Simulating birefringent noise:} We model birefringent dephasing by continuously concatenating pieces of fiber with randomly generated lengths $\triangle L$. The total propagation length thus can be split into segments of length $\Delta L$ with constant birefringence. The phase difference for the $i$-th segment is equal to the sum of $(2\pi/\lambda)\Delta L_{i} \Delta n_{i}$. These segments constitute a single phase profile associated with a particular realization of birefringent noise and corresponding changes in the refractive index difference $\triangle n$. Ensemble averaging over profiles gives density matrix for the output state depicting the random dephasing in the fiber.

\emph{Implementing DD sequences:} We investigate two DD sequences:--CPMG and XY-4 applied to the flying polarization qubits propagating through optical fiber. The basic cycles of these two sequences are $f_{\tau}Xf_{2\tau}Xf_{\tau}$ and $f_{\tau}Xf_{\tau}Yf_{\tau}Xf_{\tau}Y$, respectively, both with the cycle period of $4\tau$. The free evolution periods  correspond to the phase error accumulated during free propagation along the fiber for a length $f_{\tau}$, and the $X$ and $Y$ rotations correspond to the $\pi$ rotations implemented with half waveplates, as shown in Fig. 1. We make a comparative analysis of these sequences for the following refractive index profiles that are used to generate the errors due to the finite widths of the waveplates.
\begin{figure}[h!]
\centering
		\includegraphics[height=\columnwidth,width=2.4in,angle=-90]{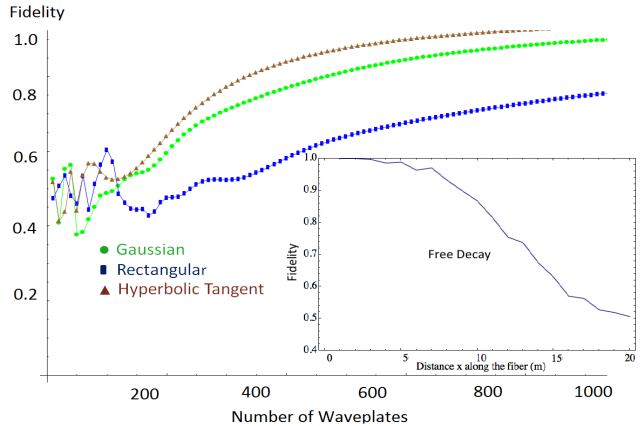}
	\caption{(Color Online) Fidelity obtained with CPMG waveplates  is shown with variation of the number of waveplates for different refractive index profiles. Total length of the optical fiber is 10 km, and number of randomly generated phase profiles to obtain the average fidelity is 500. Inset: Fidelity decay (free decay) without the waveplates in the fiber.}
\end{figure}

\emph{Generating effects of the finite width:} We aim to simulate the effect of finite widths of the waveplates when the polarization qubits are fed into the fiber. In presence of errors due to such waveplates, from Eq.~(\ref{Uc_error}) the total propagation operator after one cycle can be written as
\begin{eqnarray}
U_{C}^{i}(\tau_{p}) &=& \exp[-i \sigma_{{e}_{i}} \theta_{{e}_{i}}/2] \exp[-i \sigma_{\alpha} \theta_{p}/2] \nonumber \\
 &=&\exp[-i(\theta_{p_{i}}+\Delta \theta_{p_{i}})\sigma_{\alpha}]
\end{eqnarray}

In our model with finite-width waveplates, the angle error term $\Delta \theta_{p}$ is the practical deviation from the intended rotation, and is due to the refractive index profile $\Delta N(x)$ within the waveplates (which have width $\Delta l$). It can be written as

\begin{equation}
\Delta \theta_{p}=(2\pi/\lambda) \int_{0}^{\Delta l}\Delta N(x) dx. 
\end{equation}

We consider the following refractive index profiles for simulating realistic pulse effects:
\begin{widetext}
\begin{equation}
\label{refindex}
\Delta N(x)=\begin{cases}  \exp\left[-\frac{(x-x_{0})^{2}}{2 \sigma^{2}}\right]; 0<x_{0}<\Delta l & ~\text{(Gaussian)}\\
                       1; 0<x<\Delta l~\text{and}~0~\text{elsewhere}  & ~\text{(Rectangular)}\\
                       \tanh[a(x+1)+1]\tanh[-a(x-1)+1]; & ~\text{(Hyperbolic Tangent)}
                         \end{cases}
     \end{equation}
\end{widetext}

\begin{figure}[t!]
	\includegraphics[width=3.9in,height=0.51\textwidth,angle=-90]{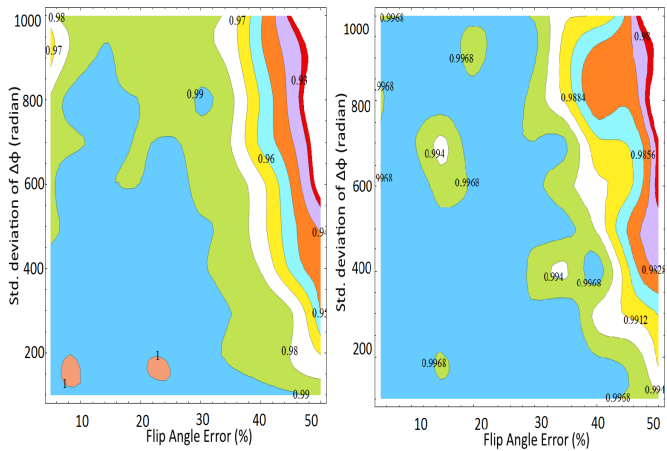}
	\caption{(Color Online) Contour plots of the fidelity with the variations of the standard deviations of  the random birefringent dephasing $\Delta \phi$ and the flip angle error for CPMG (Left) and XY-4 (Right). The simulations are done with fixed number of waveplates (1000) and total length of the fiber $L = 10$ km, and the average fidelity is obtained by taking 500 randomly generated phase profiles.}
\end{figure}

For the Gaussian profile, $x_{0}$ and $\sigma$ denote the mean and the standard deviation of the refractive index distribution within the wave plate, and for the hyperbolic tangent profile the parameter $a$ can be used to adjust the slope of the distribution inside the waveplates. In our simulation, we first consider the flip-angle error of $5\%$, which is approximately generated by using the parameters $a=8, x_{0}=1, \sigma=1.8$, from Eq.~(4.3). Refractive index profiles $\Delta N(x)$ for these parameters are shown in Fig.~2.

To characterize the effectiveness of our scheme, we use the fidelity $\mathcal{F}$ between the input state $|\psi_{\rm in}\rangle$  and $\rho_{\rm out}$ as
\begin{equation*}
\mathcal{F}=\langle\psi_{\rm in}|\rho_{\rm out}|\psi_{\rm in}\rangle,
\end{equation*}
where $\rho_{\rm out}=\frac{1}{n}$ $\displaystyle\sum\limits_{i=1}^n |\psi_{i}\rangle\langle\psi_{i}|$. Here $n$ is the total number of randomly generated phase profiles, corresponding to the propagation operator $\hat{u}_{i}$ so that $|\psi_{i}\rangle=\hat{u}_{i}|\psi_{\rm in}\rangle$ represents the simulated birefringent noise. Therefore, the fidelity being close to one implies that the input state is well-preserved against the dephasing.

The $\pi$ rotations required for CPMG and XY-4 sequences are implemented with suitably oriented half waveplates, and the effects of their finite widths on the relative random phase are generated from the refractive index profiles of Eq.~(\ref{refindex}). In Fig. 3, we show how the fidelity varies and improves with the XY-4 sequence being applied for a realistic fiber length of 10 km, even for a large variation of the parameters of the random dephasing $\Delta \phi$ for the chosen refractive index profiles. Fig. 4 illustrates the results for the CPMG sequence for the same length of the fiber. Fidelity decays without the waveplates (free decays) are also shown in the inset for comparison. In both figures, we considered 5\% flip-angle error to make the numerical results comparable.

From these figures, we find that while both the sequences work reasonably well to preserve the input polarization states for both the gaussian and hyperbolic tangent refractive index profiles, the rectangular refractive index profile gives the worst fidelity in both the cases. In fact, this profile $\emph{never}$ gives perfect fidelity with the CPMG sequence. In general, the fidelity is preserved better in case of XY-4 sequence (Fig. 3) than the CPMG (Fig. 4). This is due to the fact that the phase errors due to the finite width of the waveplates get partially cancelled due to the alternating $\pi$ rotations around two orthogonal optic axes ($x$ and $y$) in a XY-4 sequence. It is also interesting to note that the fidelity in general improves with the increasing number of pulses (waveplates) in both cases showing the robustness of these schemes in the sense that the pulse errors tend to cancel each other instead of getting added up.

The required number of waveplates to achieve a given fidelity can also be easily estimated from the above figures. For instance, for the hyperbolic tangent refractive index profiles, the required number of waveplates to achieve a $99.9\%$ fidelity, are 840 and 860 for the CPMG and XY-4 sequences, respectively. We also find that fewer number of waveplates are required for hyperbolic tangent refractive index profiles (for both the sequences) to achieve the same high fidelity, implying that our DD-sequences perform best when the finite width effects are simulated with this particular profile.

Due to the finite widths of the waveplates, the actual angle of rotation deviates from $\pi$ and this constitutes the flip angle error in the polarization state of the photon. In Fig. 5, we plot the variations of fidelity with respect to the standard deviation of the birefringent dephasing $\Delta \phi$ and flip angle errors for both the sequences. Here large flip angle errors upto 50\% are considered, and the contour plot shows that the input state is preserved upto fidelity close to one for a wide variation of the dephasing angle as well as the flip angle errors.

\section{Discussion and Conclusion}

For polarization qubit propagating through optical fibers, we demonstrate that the dephasing errors, contributed by both the fiber birefringence and the finite widths of the waveplates, could be suppressed by suitable dynamical decoupling methods. Regardless of the amplitude of the rotational error and random birefringent dephasing, our scheme provides a practical way to tackle them as long as the appropriate waveplate separations are maintained.
 As we have dealt with noises due to random fluctuations caused by any possible source such as temperature, stress, etc., the prescribed DD methods can be applied without an experimentalist having a detailed, quantitative knowledge of the decohering environment. To experimentally implement our proposed method to preserve the polarization qubits, several familiar techniques could be suitable depending on the range of fiber lengths one wishes to use. The waveplates may be directly incorporated into the fiber during the manufacturing process. Other methods include writing a Bragg transmission grating periodically into the fiber~\cite{Scalora,Hill}, or twisting the fiber in controlled ways causing suitable mechanical stress~\cite{Ulrich}. Periodic modulations or perturbations in the refractive index in the graded index optical fiber, implementing the desired profiles can be generated by the techniques described in the references~\cite{Hisatomi1,Hisatomi2,Leduc}.


We have successfully showed that it is possible to combat the random birefringent noise in an optical fiber with DD waveplates \emph{which have finite widths}. Our approach provides a practical approach to minimize errors due to both random dephasing noise in the fiber and the rotational errors due to implementation of real finite-width waveplates. This will be helpful to improve the range of quantum communication channels without requiring expensive resources such as ancilla or measurements and hence it leads to immediate commercial applications for quantum telecommunication with light. The control overhead in the proposed application of the DD sequences being reasonably small, our scheme will reduce the dephasing error while implementing a scalable quantum computing scheme with photonic qubits.

\section{Acknowledgments}

We gratefully acknowledge the support by the National Science Foundation (NSF) and the Intelligence Advanced Research Projects Activity (IARPA) via Department of Interior National Business Center contract number D12PC00527. The U.S. Government is authorized to reproduce and distribute reprints for Governmental purposes notwithstanding any copyright annotation thereon. Disclaimer: The views and conclusions contained herein are those of the authors and should not be interpreted as necessarily representing the official policies or endorsements, either expressed or implied, of IARPA, DoI/NBC, or the U.S. Government.

\bibliographystyle{h-physrev}

\end{document}